\documentclass[conference]{IEEEtran}
\IEEEoverridecommandlockouts
\usepackage{cite}
\usepackage{amsmath,amssymb,amsfonts}
\usepackage{graphicx}
\usepackage{textcomp}
\usepackage{xcolor}
\usepackage{algorithm}
\usepackage{algpseudocode}
\def\BibTeX{{\rm B\kern-.05em{\sc i\kern-.025em b}\kern-.08em
    T\kern-.1667em\lower.7ex\hbox{E}\kern-.125emX}}
\begin{document}

\title{Harnessing Supervised Learning for Adaptive Beamforming in Multibeam Satellite Systems}

\author{\IEEEauthorblockN{Flor Ortiz, Juan A. Vasquez-Peralvo, Jorge Querol, Eva Lagunas, Jorge L. Gonz\'alez Rios, Luis Garces,\\ Victor Monzon-Baeza and Symeon Chatzinotas}
\IEEEauthorblockA{\textit{Interdisciplinary Centre for Security, Reliability and Trust (SnT)}, \textit{University of Luxembourg}, Luxembourg, Luxembourg \\
Corresponding Author: flor.ortiz@uni.lu}
}

\maketitle

\begin{abstract}
In today's ever-connected world, the demand for fast and widespread connectivity is insatiable, making multibeam satellite systems an indispensable pillar of modern telecommunications infrastructure. However, the evolving communication landscape necessitates a high degree of adaptability. This adaptability is particularly crucial for beamforming, as it enables the adjustment of peak throughput and beamwidth to meet fluctuating traffic demands by varying the beamwidth, side lobe level (SLL), and effective isotropic radiated power (EIRP). This paper introduces an innovative approach rooted in supervised learning to efficiently derive the requisite beamforming matrix, aligning it with system requirements. Significantly reducing computation time, this method is uniquely tailored for real-time adaptation, enhancing the agility and responsiveness of satellite multibeam systems. Exploiting the power of supervised learning, this research enables multibeam satellites to respond quickly and intelligently to changing communication needs, ultimately ensuring uninterrupted and optimized connectivity in a dynamic world.
\end{abstract}

\begin{IEEEkeywords}
beamforming, machine learning, multibeam satellite
\end{IEEEkeywords}

\section{Introduction}
In an era defined by the insatiable demand for high-speed, ubiquitous connectivity, multibeam satellite systems have emerged as a crucial cornerstone of modern telecommunications infrastructure \cite{9049651,rs13132642}. These systems, characterized by their ability to serve multiple users and regions with a single satellite simultaneously, offer the promise of global connectivity, bridging the digital divide, and supporting a myriad of applications ranging from broadband internet access to disaster response and remote sensing \cite{giordani2020non}.

Beamforming, in essence, is the dynamic force behind the uninterrupted flow of data that connects our digital lives. Central to the performance and efficiency of multibeam satellite systems is the concept of on-board adaptive beamforming. Adaptive beamforming enables satellites to dynamically focus their transmission beams towards specific user terminals or regions, optimizing signal quality, reducing interference, and conserving precious satellite resources \cite{10197375,rs15133387}. Traditionally, adaptive beamforming algorithms have relied on fixed, pre-engineered solutions that may not fully adapt to the ever-changing conditions of the satellite environment \cite{zheng2019adaptive}. 

The paper referenced in \cite{10197375} presents an innovative approach for beam pattern synthesis tailored to the needs of geostationary satellite communication systems. This synthesis technique empowers the generation of beams characterized by a flexible beamwidth variation ranging from 0.45° to 1.5°, with independent control over these parameters for the two primary slices. The output of this advanced beam pattern synthesizer is a meticulously crafted matrix of weights imbued with beamforming coefficients precisely tailored to the desired beam characteristics. Notably, the study's results highlight the algorithm's exceptional efficacy, facilitated by the use of a surrogate optimizer. This optimizer adeptly computes the weight matrix, ultimately synthesizing beams with only minor deviations from the input data. These findings underscore the method's potential to revolutionize beam pattern synthesis for geostationary (GEO) satellite communication systems, offering a robust and responsive solution to meet the complex demands of modern telecommunications.

This paper explores a paradigm shift in the field of adaptive beamforming for multibeam satellite systems. Instead of relying solely on static beamforming techniques, we propose supervised learning for adaptive beamforming, a novel approach that harnesses the power of supervised machine learning to adaptively optimize beamforming parameters in real-time. The core of this adaptability lies within the realm of beamforming, a pivotal technology that empowers satellites to adjust their peak throughput, beamwidth, side lobe level (SLL), nulling, and effective isotropic radiated power (EIRP) to accommodate the ever-fluctuating traffic demands of our connected world \cite{10197375,8627376}.

The main idea behind SLAB is to equip multibeam satellite systems with the ability to autonomously learn and adapt their beamforming strategies based on historical data and ongoing environmental conditions \cite{8627376,8929162}. By doing so, supervised learning for adaptive beamforming promises to revolutionize the field, enabling satellites to operate more efficiently, reduce latency, enhance data rates, and extend the reach of their services \cite{aerospace10020101}.

We present the theoretical underpinnings of supervised learning for on-board adaptive beamforming, provide a comprehensive overview of the supervised learning techniques employed, and detail the implementation of this innovative approach within multibeam satellite systems. We also offer an extensive performance evaluation, comparing our approach to traditional beamforming methods under various operational scenarios and illustrating the substantial benefits it provides.

\section{System Model and Problem Definition}

In the context of designing a multibeam satellite antenna system, several key factors need to be considered, including antenna array dimensioning, beamwidth, SLL, nulling, and EIRP control. 

\subsection{Antenna Array Dimensioning}
This work considers a direct-radiating antenna array (DRA).
The number of antenna elements in the DRA is determined by the required gain, beam solid angle, satellite altitude, position, and coverage area. For our study, we assume that the satellite is positioned in the geostationary orbit. Assuming a symmetric and rectangular array comprising $N\times N$ radiating elements, the size of each dimension can be calculated as:
\begin{equation}
    N = \frac{\mathrm{asinc}(\frac{1}{\sqrt2})\lambda_0}{\eta\ \theta_{-3dB}  2 d},
\end{equation} 
where, $d$ represents the inter-element spacing, $\lambda_0$ is the operating wavelength, $\theta_{-3dB}$ is the beamwidth, and $\eta$ is the antenna efficiency. For our scenario, we employ open-ended waveguide antennas as the radiating elements, with an inter-element separation of 7/8$\lambda_0$ chosen to maintain mutual coupling below -30~dB for a central frequency $f_0$ = 19~GHz \cite{10197375}. The estimated efficiency is approximately 90\% to account for potential factors affecting overall performance. Thus, the proposed DRA has a total of $144\times144$ antenna elements.

\begin{figure*}[!h]
         \centering
        	\includegraphics[width=18.cm]{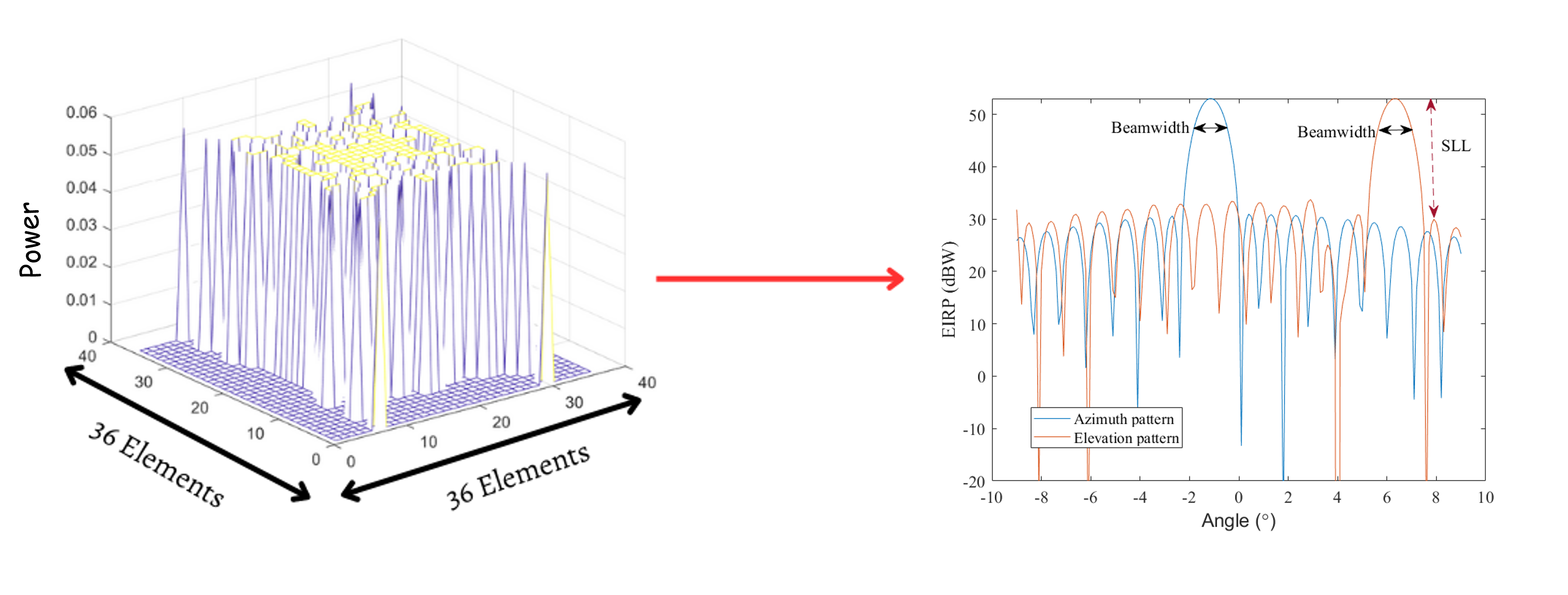}
        \caption{Proposed DRA: $36\times36$ sub-arrays (elements), where each element can be active or not. Each  element weight is defined per beam and defines the beamwidth in $b-th$ beam, $\theta_t^b$, an $EIRP_t^b$, and a specific SLL}
        \label{fig:BMatix}
    \end{figure*}

Individually controlling all the radiating elements in this array would be impractical due to the high number of required radio-frequency (RF) chains (one per antenna).
To mitigate this, we partition the array in 4$\times$4 antennas sub-arrays, which results in a new unit cell dimension and an inter-element spacing of 3.5$\lambda_0$ \cite{10197375}.
Consequently, the number of RF chains reduces to 36$\times$36, each connected to one of the sub-arrays or unit cells (referred to as elements in the rest of the paper). 
This configuration of the beamforming matrix will be reflected in the radiation pattern provided by each beam, where we will obtain a beamwidth in $b-th$ beam, $\theta_t^b$, an $EIRP_t^b$, and a specific SLL (see Fig. \ref{fig:BMatix}). 

\subsection{Beamwidth, SLL, Nulling, and EIRP Control}

To optimize the antenna system, several parameters require control:
\begin{itemize}
    \item \textbf{Scanning Angles:} Steering the beam is accomplished by modifying the complex component of the weight matrix. This can be achieved through progressive phase shifts, FFT-based methods \cite{Ha2022,Palisetty2022}, or codebook-based beamforming \cite{5455331}. We calculate incremental phase shifts using 
\begin{equation}
    \label{Eq.incrementalPhase}
    \Theta_{\mathrm{mn}} = k(md_x\sin(\theta_0)\cos(\phi_0)+nd_y\sin(\theta_0)\sin(\phi_0)),
\end{equation}
where $k=\frac{2\pi}{\lambda_0}$ is the wave number, $m$ and $n$ are the positions of the elements in the $x$ and $y$-axis, respectively, $dx$ and $dy$ are the corresponding periods, and $\theta_0$ and $\phi_0$ are the scanning angles \cite{5455331}.
\item \textbf{Beamwidth and SLL Control:} Beamwidth and SLL are controlled using tapering techniques. Chebyshev amplitude tapering is chosen due to its effectiveness in narrowing the beam \cite{1139022}.
\item \textbf{Nulling Control:} Nulling is achieved by modifying the antenna's progressive phase shift. $W_{\theta_0\phi_0}$ represents the weight matrix with the progressive phase shift to the steering direction, and $W_{\mathrm{null}}$ represents the weight matrix with the progressive phase shift towards the desired nulling angle. 
\item \textbf{EIRP Control:} EIRP control depends on beamwidth, scanning angle, and power radiated by each antenna element. Compensating for scanning losses and achieving the desired EIRP may require adjusting the power per element, especially in scenarios involving subarrays.
\end{itemize}

\subsection{Beamforming Cost Function}
To encapsulate these considerations, we define a beamforming cost function with the objective of minimizing $\mathrm{F}(Z_1+Z_2+Z_3)$. This cost function comprises three sub-objectives, each addressing different aspects of beamforming optimization.

The first sub-objective quantifies the error between the desired beamwidths, both in azimuth ($\mathrm{\theta_{-3dB_{Az_o}^b}}$ and $\mathrm{\theta_{-3dB_{Az_c}^b}}$) and elevation ($\mathrm{\theta_{-3dB_{El_o}^b}}$ and $\mathrm{\theta_{-3dB_{El_c}^b}}$), for each beam. 

The second sub-objective assesses the error between the minimum SLL requirements, in both azimuth ($\mathrm{SLL_{Az_o}^b}$ and $\mathrm{SLL_{Az_c}^b}$) and elevation ($\mathrm{SLL_{El_o}^b}$ and $\mathrm{SLL_{El_c}^b}$), and the SLL achieved by the beamforming matrix.

Lastly, the third sub-objective calculates the error between the desired EIRP ($\mathrm{EIRP_o^b}$) and the computed EIRP ($\mathrm{EIRP_c^b}$) for each beam.

These terms are weighted by factors $k_1$, $k_2$, and $k_3$, allowing for fine-tuned adjustments to their importance within the optimization process.

The optimization problem can be succinctly expressed as follows:
\begin{equation}
\label{eq:CostFunction}
\min_{W_{p\times q}^B} \left( Z_1(W_{p\times q}^B) + Z_2(W_{p\times q}^B) + Z_3(W_{p\times q}^B) \right),
\end{equation}
where:
\begin{equation*}
\label{eq:Eq2bF}
    \left\{
    \begin{aligned}
        & Z_1 =  \Bigg( \frac{|\mathrm{\theta_{-3dB_{Az_c}}^b}(W_{p\times q}^B)-\mathrm{\theta_{-3dB_{Az_o}}^b}|}{\mathrm{\theta_{-3dB_{Az_o}}^b}}+ \\ 
        & \hspace{1cm} \frac{|\mathrm{\theta_{-3dB_{El_c}}^b}(W_{p\times q}^B)-\mathrm{\theta_{-3dB_{El_o}}^b}|}{\mathrm{\theta_{-3dB_{El_o}}^b}} \Bigg)k_1\\  
        & Z_2 = \Bigg(\frac{|\mathrm{SLL_{Az_c}^b}(W_{p\times q}^B)-\mathrm{SLL_{Az_o}^b}|}{\mathrm{SLL{Az_o}^b}}+ \\
        & \hspace{1cm} \frac{|\mathrm{SLL_{El_c}^b}(W_{p\times q}^B)-\mathrm{SLL_{El_o}^b}|}{\mathrm{SLL{El_o}^b}}\Bigg)k_2\\
        &  Z_3 = \left(\frac{\mathrm{EIRP_c^b}(W_{p\times q}^B)-\mathrm{EIRP_o^b}}{\mathrm{EIRP_o^b}}\right)k_3. \\
    \end{aligned}
    \right.
\end{equation*}

\section{Supervised Learning for Adaptive Beamforming}
In this section, we present our classification approach for selecting the optimal beamforming matrix using neural networks. We aim to employ a neural network-based classifier to predict the most suitable beamforming matrix for a given set of input parameters.

\subsection{Beamforming Matrix Clustering}
    \begin{figure*}[!h]
         \centering
        	\includegraphics[width=16.cm]{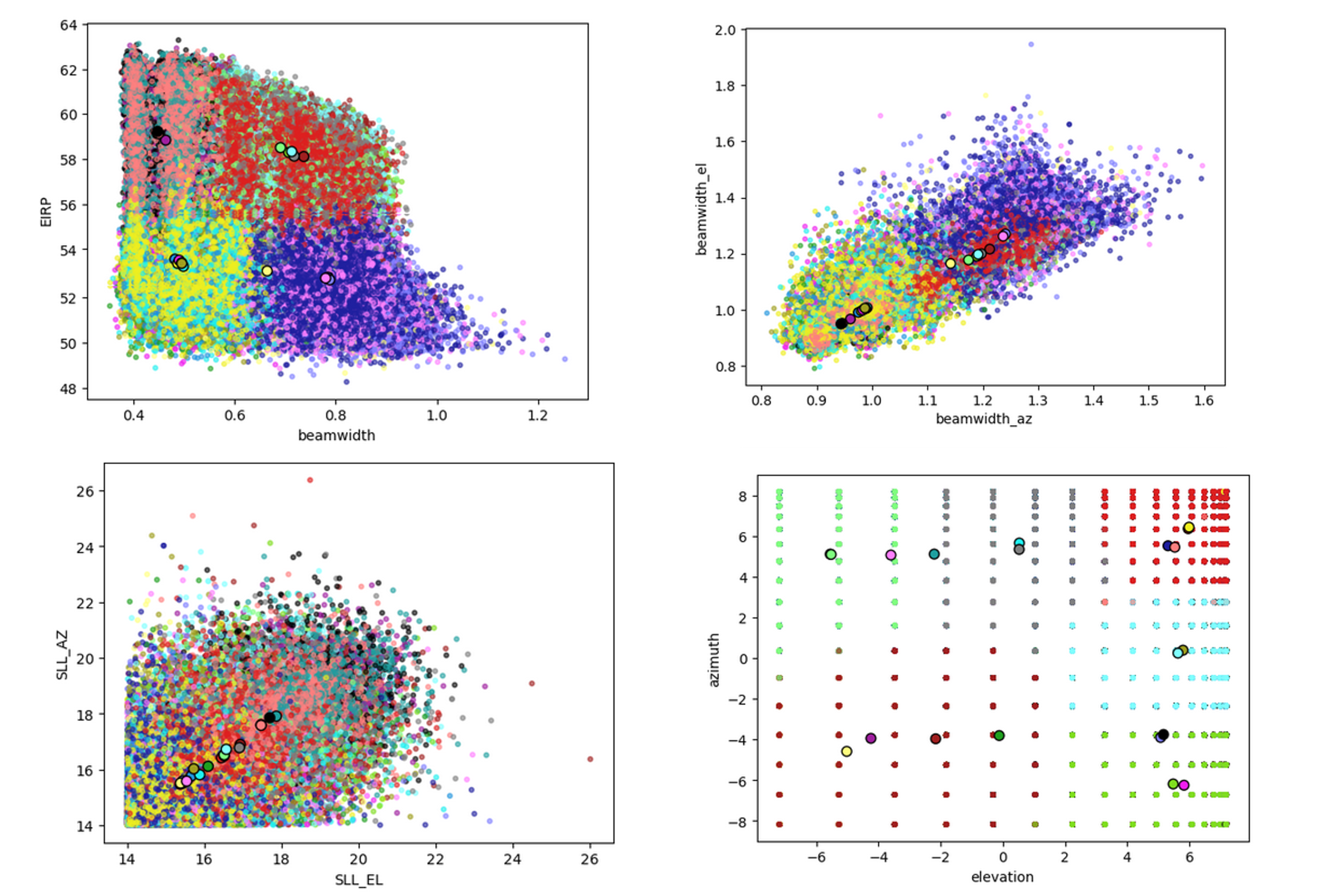}
        \caption{Analysis of key input variables that affect beamforming design. Each color represents one cluster}
        \label{fig:cluste}
    \end{figure*}
    
The core challenge we confront is the vast search space of potential beamforming matrices. As elucidated in Section II, our beamforming matrix consists of 36x36 elements, totaling 1296 individual components. Each element can either be activated or remain inactive, resulting in an astronomically large number of possible combinations. Effectively navigating this expansive solution space necessitates a method that can distill and comprehend the intricate relationships between various parameters, such as azimuth and elevation beamwidths ($\theta_{3dB,ele}$ and $\theta_{3dB,azi}$), SLL in elevation and azimuth ($SLL_{el}$ and $SLL_{az}$), desired EIRP, and pointing coordinates in elevation and azimuth.

To tackle this complexity, we employ a clustering-based approach, specifically leveraging the K-means algorithm. This algorithm plays a pivotal role in grouping and categorizing influential variables within our system, ultimately shedding light on their combined impact on the design of the beamforming array. The variables considered for clustering include $\theta_{3dB,ele}$ (elevation beamwidth), $\theta_{3dB,azi}$ (azimuth beamwidth), $SLL_{el}$ (SLL in elevation), $SLL_{az}$ (SLL in azimuth), $EIRP$, elevation, and azimuth (as illustrated in Figure \ref{fig:cluste}).

The K-means algorithm aims to minimize the within-cluster sum of squares, which can be formulated as follows:

\begin{equation}
\label{eq:kmeans-objective}
\min_{\substack{S}} \sum_{i=1}^{K} \sum_{\mathbf{x} \in S_i} ||\mathbf{x} - \boldsymbol{\mu}_i||^2,
\end{equation}

where $K$ represents the number of clusters, $S_i$ is the set of data points assigned to cluster $i$, $\mathbf{x}$ represents a data point,
$\boldsymbol{\mu}_i$ is the centroid of cluster $i$.

Utilizing clustering, we identify similar sets of input parameters and assign them to corresponding clusters. Each cluster is associated with a specific pre-stacked beamforming matrix, effectively encapsulating a subset of the vast dataset. This transformative approach effectively converts the problem into a binary classification scenario, where each class corresponds to a distinct pre-defined matrix within our extensive set of beamforming options.

Using $K$ clusters as representative classes, each aligned with a specific set of input data points and their corresponding beamforming matrices. This approach forms the foundation of our supervised learning framework, facilitating the intelligent adaptation of beamforming in response to dynamic communication requirements.

\subsection{Classification Approach to select the best Beamforming Matrix}


    \begin{figure}[!h]
         \centering
        	\includegraphics[width=\columnwidth]{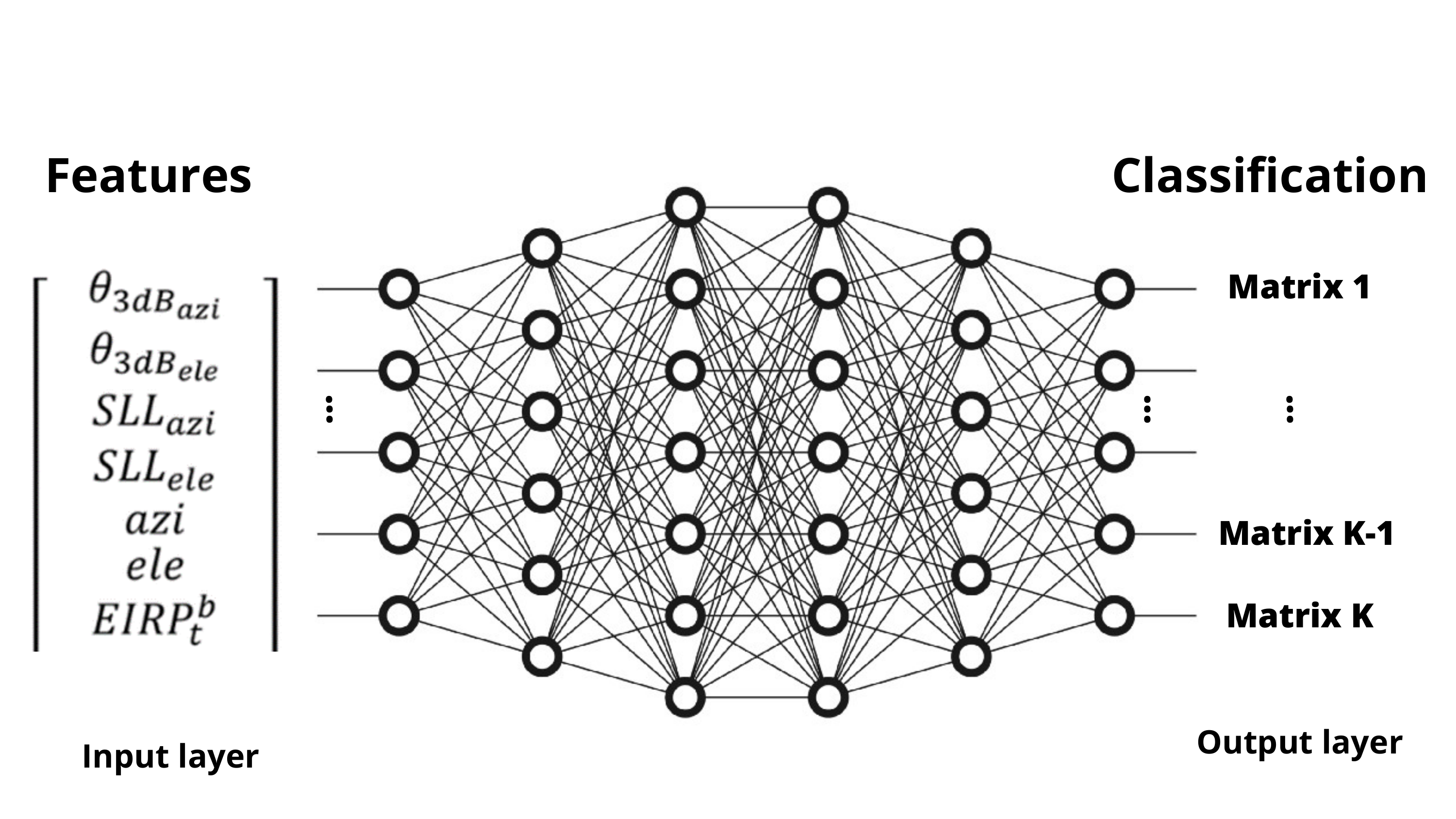}
        \caption{Artificial Neural Network for Classification Approach to select the best Beamforming Matrix}
        \label{fig:ANN}
    \end{figure}

\begin{algorithm}[h]
\caption{Beamforming Matrix Selection using Neural Networks}
\begin{algorithmic}[1]
\Procedure{TrainNeuralNetwork}{$\mathbf{X}$, $\mathbf{Y}$}
\State Initialize neural network architecture
\State Define loss function $L$ (Categorical Cross-Entropy)
\State Initialize optimizer (e.g., SGD, Adam)
\While{Not converged}
    \State Forward pass: Compute predicted probabilities
    \State Calculate loss using Equation (\ref{eq:cross-entropy-loss2})
    \State Backward pass: Compute gradients
    \State Update neural network weights using optimizer
\EndWhile
\EndProcedure

\Procedure{SelectBeamformingMatrix}{$\mathbf{X_{new}}$}
\State Load pre-trained neural network
\State Perform forward pass with $\mathbf{X_{new}}$
\State Get cluster probabilities for each class
\State Select cluster with highest probability as beamforming matrix
\Return Selected beamforming matrix
\EndProcedure
\end{algorithmic}
\end{algorithm}

Our classification model is based on a feedforward neural network. Let $\mathbf{X}$ represent the input feature vector, which comprises parameters such as azimuth and elevation beamwidths ($\theta_{3dB,ele}$ and $\theta_{3dB,azi}$), Side Lobe Level in elevation and azimuth ($SLL_{el}$ and $SLL_{az}$), desired $EIRP$, $elevation$, and $azimuth$. The output layer of the neural network consists of $K$ neurons, each corresponding to one of the $K$ pre-defined clusters of beamforming matrices.

The neural network's architecture can be summarized as follows (See Fig. \ref{fig:ANN}):
\begin{enumerate}
    \item \textbf{Input Layer:} The input layer consists of neurons representing the input features $\mathbf{X}$.
    \item \textbf{Hidden layers:} We incorporate one or more hidden layers to capture complex relationships within the data. Each hidden layer contains a varying number of neurons. The choice of the number of hidden layers and neurons per layer can be determined through experimentation.
    \item \textbf{Output Layer:} The output layer comprises $K$ neurons, where $K$ represents the number of clusters determined by the K-means algorithm in the "Beamforming Matrix Clustering" approach. Each output neuron corresponds to one of the $K$ pre-defined clusters.
\end{enumerate}

The neural network training involves supervised learning, where we use labeled data to teach the model to predict the appropriate beamforming matrix cluster. The loss function used for training can be a categorical cross-entropy loss, defined as \cite{https://doi.org/10.1002/sat.1422,cunningham2008supervised}:

\begin{equation}
\label{eq:cross-entropy-loss}
L(\mathbf{X}, y) = -\sum_{i=1}^{K} y_i \log(p_i),
\end{equation}

where, $y_i$ is the ground truth label for class $i$ (one-hot encoded), $p_i$ is the predicted probability of the input belonging to class $i$.

For a dataset with $S$ samples in all, the categorical cross-entropy loss is given by \cite{cornejo2022method}:

\begin{equation}
\label{eq:cross-entropy-loss2}
L(\mathbf{X}, y) = -\sum_{j=1}^{S}\sum_{i=1}^{K} y_i \log(p_i),
\end{equation}

The neural network is trained to minimize this loss function using optimization techniques such as stochastic gradient descent (SGD) or Adam optimization.

Once the neural network is trained, it can be used for inference to select the most appropriate beamforming matrix for a given set of input parameters. The neural network predicts the probability distribution over the $K$ clusters when new input data is provided. The cluster with the highest predicted probability is selected as the best beamforming matrix.

Algorithm 1 shows a simplified pseudocode for the classification approach using neural networks to select the best beamforming matrix.

\section{Numerical Results}
To facilitate the mapping of input data to the most suitable beamforming matrix, we chose to use 20 clusters as representative classes. Each class corresponds to a pre-defined matrix within the set of 20 matrices. 

Categorizing the input data into these distinct classes establishes a clear mapping between each data point and the corresponding pre-defined matrix. This mapping enables us to effectively assign and utilize the appropriate beamforming matrix based on the input variables' characteristics during the beamforming process.

        \begin{figure}[!h]
         \centering
        	\includegraphics[width=8.cm]{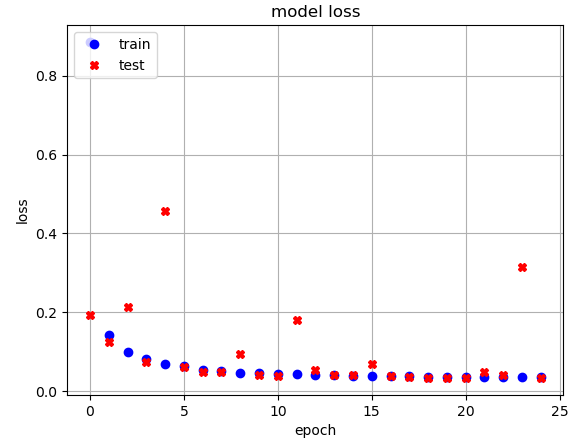}
        \caption{Loss performance during the training for training and validation data}
        \label{fig:Loss2}
    \end{figure}

        \begin{figure}[!h]
         \centering
        	\includegraphics[width=8.cm]{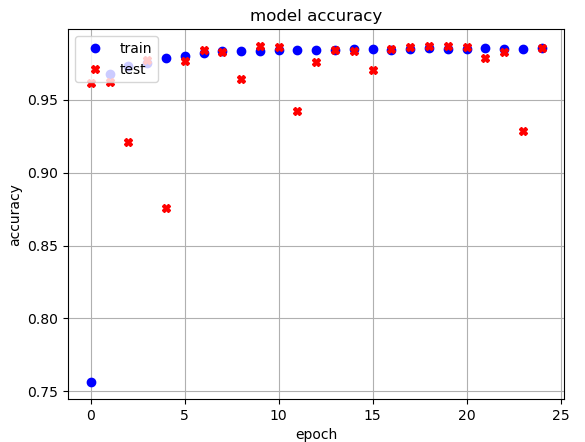}
        \caption{$Accuracy$ performance during the training for training and validation data}
        \label{fig:ACC2}
    \end{figure}

We evaluate the performance of our model using key metrics, including loss and accuracy. Figures \ref{fig:Loss2} and \ref{fig:ACC2} display the training and validation results, showcasing the model's effectiveness. We achieved a training and validation loss of less than 0.03, indicating that the model effectively learned to classify input data into the predefined classes. Our model achieved an accuracy greater than $97\%$ in both training and validation, affirming its ability to accurately predict the appropriate beamforming matrix.

\begin{figure}[!h]
         \centering
        	\includegraphics[width=8.cm]{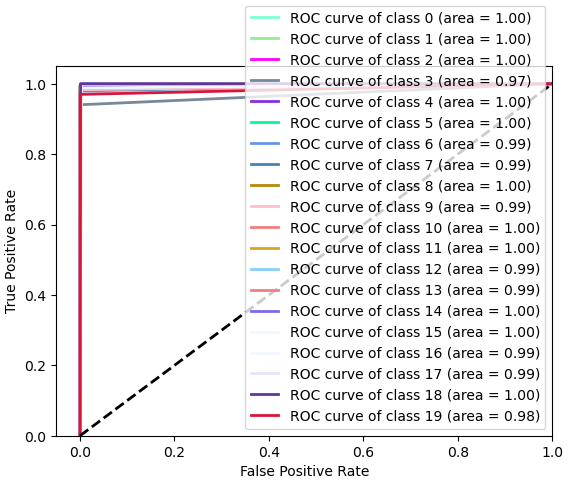}
        \caption{Confusion Matrix for approach 2}
        \label{fig:ROC_Beam}
    \end{figure}

Figure \ref{fig:ROC_Beam} shows the ROC curve that illustrates the trade-off between sensitivity (true positive rate) and specificity (true negative rate) for different classification thresholds. Although our classification problem involves 20 classes, the ROC curve is generated by considering each class against the rest, offering insights into the model's discrimination ability. The ROC curve demonstrates the model's precision, with an accuracy exceeding $94\%$ for each class.
    
\begin{figure}[!h]
         \centering
        	\includegraphics[width=10.cm]{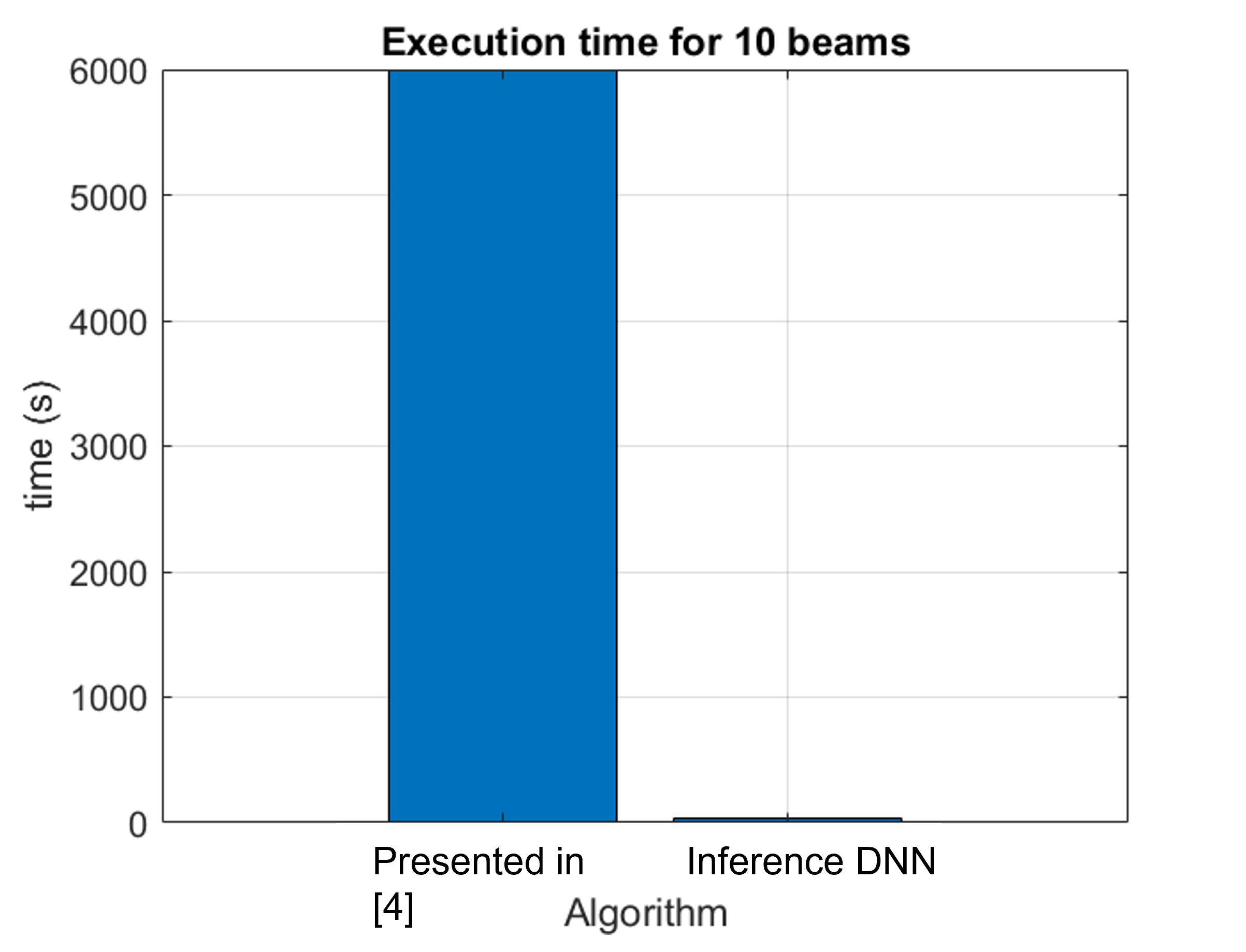}
        \caption{Run time to obtain the Beamforming Matrix for 10 beams using Algorithm in \cite{10197375} and Inference using DNN}
        \label{fig:time}
    \end{figure}

In comparison with the algorithm presented in \cite{10197375}, our approach offers significant advantages in terms of execution time. While the algorithm in \cite{10197375} requires at least 10 minutes to obtain a beamforming matrix for each beam, our deep neural network (DNN) model, once trained, takes only about 3 seconds. Figure \ref{fig:time} provides a comprehensive time comparison for a system with 10 beams, highlighting the efficiency of our approach.

\begin{figure}[!h]
         \centering
        	\includegraphics[width=8.cm]{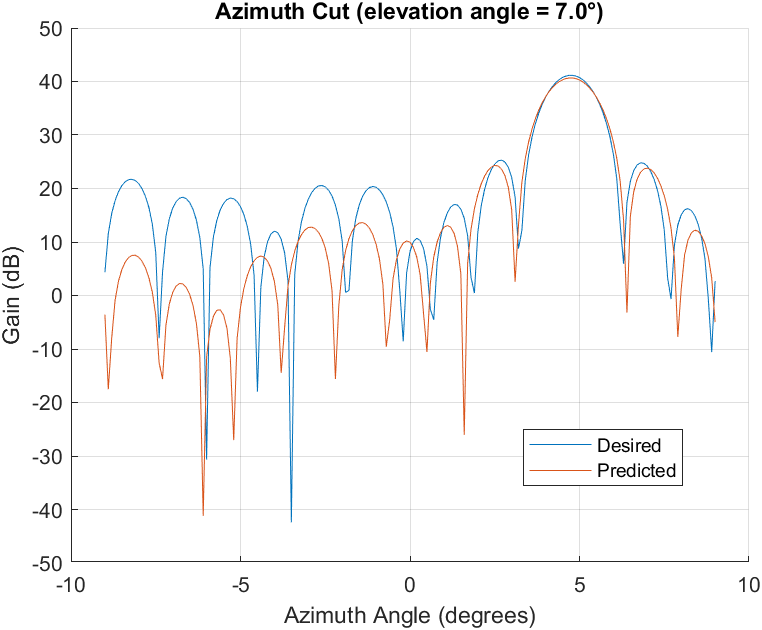}
        \caption{Radiation Pattern. Azimuth Cut}
        \label{fig:azim}
    \end{figure}

\begin{figure}[!h]
         \centering
        	\includegraphics[width=8.cm]{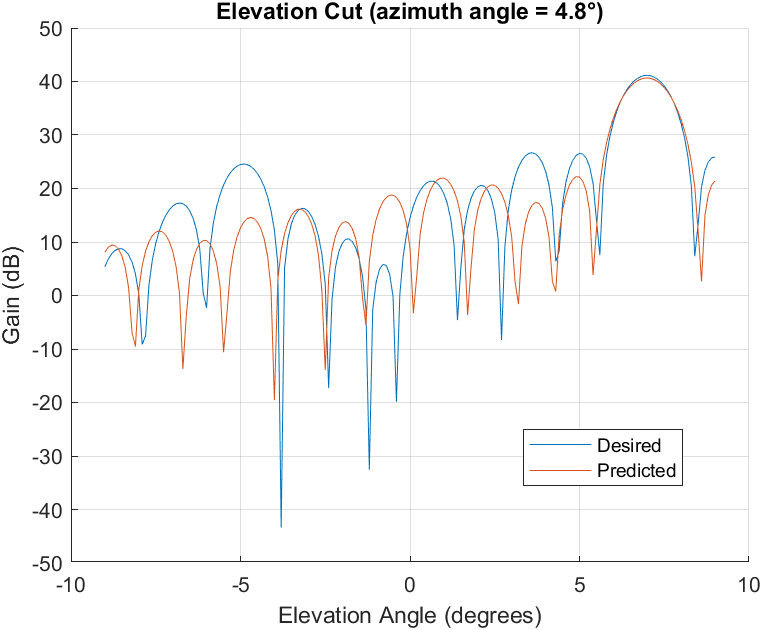}
        \caption{Radiation Pattern. Elevation Cut}
        \label{fig:elev}
    \end{figure}

Figures \ref{fig:azim} and \ref{fig:elev} depict the radiation patterns obtained through our model's predictions for azimuth and elevation. These patterns align with the desired specifications, meeting the minimum SLL requirements and exhibiting beamwidth differences of only 0.03 degrees. Additionally, the gain differences are less than 0.5 dB, highlighting the model's ability to optimize system performance.

\section{Conclusion}
Our model exhibited remarkable performance. Comparing our approach to the algorithm presented in \cite{10197375}, we significantly reduced execution time, making real-time beamforming feasible.

Moreover, our model met and improved system requirements, ensuring minimum side lobe levels and precise beamwidth control. For future work, we plan to explore the scalability of our approach to handle a larger number of beams and classes effectively. Additionally, we aim to enhance the model's adaptability to dynamic communication environments, allowing it to respond dynamically to changing requirements. Integrating reinforcement learning techniques may further optimize beamforming decisions for evolving satellite communication needs \cite{mismar2019deep}.

\section*{Acknowledgment}
This work was supported by the European Space Agency (ESA) funded under Contract No. 4000134522/21/NL/FGL named “Satellite Signal Processing Techniques using a Commercial Off-The-Shelf AI Chipset (SPAICE)”. Please note that the views of the authors of this paper do not necessarily reflect the views of the ESA. Furthermore, this work was partially supported by the Luxembourg National Research Fund (FNR) under the project SmartSpace (C21/IS/16193290). 


\bibliographystyle{IEEEtran}
\bibliography{references}

\end{document}